\documentclass[pra,showpacs,12pt,tightenlines]{revtex4}
\usepackage{amsmath}
\usepackage{amssymb}
\usepackage{dsfont}
\usepackage{graphicx}
\newcommand{\half}{\mbox{$\textstyle \frac{1}{2}$}}
\newcommand{\quat}{\mbox{$\textstyle \frac{1}{4}$}}

\newcommand{\ri}{\mbox{$\rm i$}}
\newcommand{\rd}{\mbox{$\rm d$}}

\begin{document}
\title[information geometry and density matrices]
{Information geometry of density matrices and state estimation}
\author{Dorje C. Brody}
\affiliation{Mathematical Sciences, Brunel University, Uxbridge UB8 3PH, UK} 

\begin{abstract}
Given a pure state vector $|x\rangle$ and a density matrix ${\hat\rho}$, 
the function $p(x|{\hat\rho})=\langle x|{\hat\rho}|x\rangle$ defines a 
probability density on the space of pure states parameterised by density 
matrices. The associated Fisher-Rao information measure is used to 
define a unitary invariant Riemannian metric on the space of density 
matrices. An alternative derivation of the metric, based on 
square-root density matrices and trace norms, is provided. This 
is applied to the problem of quantum-state estimation. In the simplest 
case of unitary parameter estimation, new higher-order corrections to the 
uncertainty relations, applicable to general mixed states, are derived. 
\end{abstract}

\maketitle
\vspace{0.4cm}



The structure of the space ${\mathcal D}$ of density matrices is 
of considerable interest in quantum 
information theory and related disciplines such as quantum 
tomography and quantum statistical inference. While substantial 
progress has been made in unravelling the structure of ${\mathcal 
D}$ \cite{Hubner,BZ,Petz,Uhlmann,Hansen}, the subject remains difficult, 
partly because of the complicated structure of ${\mathcal D}$, 
which is a manifold with a boundary 
given by degenerate density matrices. 
The purpose of the present paper 
is to reveal the existence of a remarkably simple dual metric 
structure over the space of pure and mixed states, and to apply 
this to obtain new higher-order corrections to Heisenberg 
uncertainty relations for general mixed states. 

The idea of the dual metric structure can be sketched as follows. 
If $|x\rangle$ denotes a normalised 
pure state vector and ${\hat\rho}$ a generic density matrix, then the 
expectation $\langle x|{\hat\rho}|x\rangle$ defines a probability 
density function either on the space of pure states parameterised 
by density matrices, or on the space of density matrices 
parameterised by pure states. Hence, methods of information 
geometry \cite{Rao,Amari,Brody0} 
can be employed to define a pair of Riemannian metrics, one for 
the pure state space and one for the space of density matrices. 

An alternative formulation is to use the trace norm and matrix 
algebra associated with the Hermitian square-root map ${\hat\rho} \to 
{\hat\xi} = \surd{\hat\rho}$. It is shown that if ${\hat\xi}$ is 
parameterised by a pure state so that ${\hat\xi}={\hat\rho}$, 
then the associated information metric is the Fubini-Study 
metric; whereas if ${\hat\xi}$ is parameterised by a generic 
density matrix, then we recover the information metric on 
${\mathcal D}$. Apart from these two extremes, however, 
there are many other cases of considerable interest. 
Typically, in a problem of quantum state estimation, the density 
matrix ${\hat\rho}(\theta)$ depends on one 
or several unknown parameters $\{\theta^i\}$ that are to be estimated. 
The Fisher-Rao information metric $4\,{\rm tr}[\partial_i 
{\hat\xi}(\theta)\partial_j {\hat\xi}(\theta)]$, where $\partial_i = 
\partial/\partial\theta^i$, then determines the covariance lower bound 
of the parameter estimates. In the simplest case in which 
${\hat\rho}(\theta)$ represents a one-parameter unitary curve 
in the space of density matrices (e.g., $\theta$ represents 
time), the quantum Fisher-Rao metric coincides with the ``skew 
information'' defined by Wigner and Yanase \cite{WY}. By this 
device we are able to extend the results of 
\cite{BC,FN,caianiello,Brody1,Luo,Luo3} 
on quantum statistical inference. In particular, we derive: 
(a) a variance lower bound for general mixed states, which is sharper 
than the previously known ones that are expressed in terms of the 
skew information; (b) the quantum-mechanical analogue of the 
conditional variance representation in terms of the skew information 
measures; and (c) higher-order corrections to the uncertainty lower bound 
for mixed states, whereby the concept of higher-order skew information 
measures are introduced. 

Consider a finite, $n$-dimensional quantum system. The space 
of pure states ${\mathit\Gamma}$ is the space of rays through the 
origin of ${\mathds C}(n)$, which is just the complex projective 
space equipped with the unitary 
invariant Fubini-Study metric. We let ${\mathcal D}$ denote the 
space of density matrices on ${\mathds C}(n)$. Every density 
matrix defines a positive scalar function 
$\langle x|{\hat\rho}|x\rangle$ on 
${\mathit\Gamma}$, and this, properly normalised with respect to 
the Fubini-Study volume element, yields a probability density on 
${\mathit\Gamma}$. Since, by the real polarisation identity, a 
density matrix is uniquely determined by the totality of its 
expectation values, these probability distributions are parameterised 
by ${\mathcal D}$ in a one-to-one fashion. Hence we can use the 
Hellinger distances between these distributions and obtain an 
\textit{information metric} on ${\mathcal D}$ to investigate properties 
of ${\mathcal D}$, which are directly related with the physical 
significance of density matrices. 
The dual picture emerges from the converse construction: Given any 
$|x\rangle\in{\mathit\Gamma}$, one obtains a nonzero nonnegative 
function $\langle x|{\hat\rho}|x\rangle$ 
on ${\mathcal D}$, which defines, when properly normalised, a set of 
probability distributions on ${\mathcal D}$. This
likewise yields an information metric on ${\mathit\Gamma}$. 
By considerations of symmetry and invariance, using the unitary
structure of ${\mathds C}^n$, one can
perceive that this \textit{a priori} new metric on
${\mathit\Gamma}$ can only be the Fubini-Study metric.

If one assumes the Schr\"odinger equations of motion 
$|x(t)\rangle = \exp(-\ri{\hat H}t)|x(0)\rangle$ on 
${\mathit\Gamma}$ and the Heisenberg equations of motion 
${\hat\rho}(t) = \exp(-\ri{\hat H}t) {\hat\rho}(0)\exp(\ri{\hat H}t)$ 
on ${\mathcal D}$, then the following conclusions can be drawn. 
The evolution on ${\mathit\Gamma}$ is isometric with respect to 
the Fubini-Study metric (known as Wigner's theorem), and, 
moreover, the evolution on ${\mathcal D}$ is isometric with 
respect to the information metric. Thus, we have a complete 
reciprocity between the Schr\"odinger dynamics on 
${\mathit\Gamma}$, equipped with the Fubini-Study metric, 
and the Heisenberg dynamics on ${\mathcal D}$, endowed 
with the information metric. 

The components of these information metrics, with respect to a 
given choice of coordinate system, can be calculated by use of 
the standard expression for the Fisher-Rao metric. Let us write 
$\{x^\mu\}_{\mu=1,\ldots,2n-2}$ for the coordinates on 
${\mathit\Gamma}$ 
with volume element $\rd v_x$, and $\{\rho^a\}_{a=1,\ldots,n^2-1}$ for 
the coordinates on ${\mathcal D}$ with volume element $\rd V_\rho$. 
Likewise, we write $p(x|{\hat\rho})\propto\langle x|
{\hat\rho}|x\rangle$ for the normalised density on 
${\mathit\Gamma}$ and $P({\hat\rho}|x)\propto\langle x|
{\hat\rho}|x\rangle$ for the normalised density on ${\mathcal D}$. 
Then the information metric $G_{ab}$ on ${\mathcal D}$ is given 
by 
\begin{eqnarray}
G_{ab}({\hat\rho}) = \int_{\mathit\Gamma}  \frac{1}{p(x|{\hat\rho})} 
\frac{\partial p(x|{\hat\rho})}{\partial\rho^a}
\frac{\partial p(x|{\hat\rho})}{\partial\rho^b} \, \rd v_x. \label{eq:1}
\end{eqnarray}
Similarly, the information (Fubini-Study) metric $g_{\mu\nu}$ on 
${\mathit\Gamma}$ is given by 
\begin{eqnarray}
g_{\mu\nu}(x) = \int_{\mathcal D} \frac{1}{P({\hat\rho}|x)} 
\frac{\partial P({\hat\rho}|x)}{\partial x^\mu}
\frac{\partial P({\hat\rho}|x)}{\partial x^\nu} \, \rd V_\rho . \label{eq:2}
\end{eqnarray}
We therefore have two statistical manifolds, ${\mathit\Gamma}$ 
with the metric $g_{\mu\nu}$, and 
${\mathcal D}$ with the metric $G_{ab}$; 
each of these is represented by probability distributions on the 
other. 

The duality of information metrics leads to integral representations  
for observable expectations. Let us consider trace-free 
observables ${\hat A}$. Defining 
$A(x) = \langle x|{\hat A}|x\rangle$ and $A({\hat\rho}) = 
{\rm tr}({\hat\rho}{\hat A})$, the dual formulae read 
\begin{eqnarray}
A({\hat\rho}) = \int_{\mathit\Gamma} A(x) p(x|{\hat\rho})\, {\rd}v_x 
\quad {\rm and} \quad 
A(x) = \int_{{\mathcal D}} A({\hat\rho}) P({\hat\rho}|x) {\rd}V_{\rho}.
\end{eqnarray}
The first formula is a special case of Gibbons' explicit formula
for the trace of a product of two observables \cite{Gibbons}. The 
second is proved using Gibbons' formula along with  
invariance arguments. 

We remark that the space of density matrices ${\mathcal D}$ is a 
manifold with a 
piecewise smooth boundary, the latter consisting of the degenerate 
density matrices, i.e. those with at least one zero eigenvalue. The 
structure of ${\mathcal D}$ is thus analogous to that of the unit 
$n-1$ simplex, which is the space of probability distributions on a 
set of $n$ points. Information geometry is indeed applicable, 
\textit{mutatis mutandis}, to manifolds with boundary. Thus, 
defining the information metric on ${\mathcal D}$ in the above 
described manner, one obtains a volume element with the necessary 
invariance properties to prove 
the dual integral formulae. In particular, the Heisenberg dynamics 
on ${\mathcal D}$ defines a Killing field with respect to this metric, 
which leaves the set of degenerate 
matrices (i.e. the boundary) invariant, that is, the Killing field is 
tangent to the boundary. 

An alternative approach is to label density matrices by their Hermitian 
square roots. The significance of the square-root map in the context of 
classical probability is outlined in Rao's seminal paper \cite{Rao}; also 
elaborated in \cite{Brody0}. The key formulation here  in the quantum 
context is to consider not arbitrary square root of the density matrix, but 
the Hermitian square root, which provides the natural extension of the 
classical theory. 
Thus, we write ${\hat\rho}={\hat\xi}{\hat\xi}$, 
where ${\hat\xi}$ is an arbitrary Hermitian matrix such that 
${\rm tr}({\hat\xi}^2)=1$. If ${\hat\rho}$ is a pure state so that 
${\hat\rho}={\hat\xi}=|x\rangle\langle x|$, then the quantum 
Fisher-Rao information metric is given by $g_{\mu\nu}=4{\rm tr} 
(\partial_\mu {\hat\xi} \partial_\nu {\hat\xi})$, where $\partial_\mu = 
\partial/\partial x^\mu$, which is just the Fubini-Study 
metric (\ref{eq:2}). On the other hand, if ${\hat\rho}$ is a generic 
mixed state the quantum Fisher-Rao metric is $G_{ab}=4{\rm tr} 
(\partial_a {\hat\xi} \partial_b {\hat\xi})$, where $\partial_a = 
\partial/\partial \rho^a$, and this is just the information metric 
(\ref{eq:1}). 

As an illustration, consider the $2\times2$ case. A  generic 
Hermitian matrix ${\hat\xi}$ can be expressed in the form  
\begin{eqnarray}
{\hat \xi} = \left( \begin{array}{cc} t+z & x-{\rm i}y \\
x+{\rm i}y & t-z \end{array} \right) . \label{eq:5}
\end{eqnarray}
For a pure state we write $(x^1,x^2)=(\theta,\phi)$, where 
$t=\half$, $x=\half \sin\theta\cos\phi$,  $y=\half \sin\theta \sin\phi$, 
and  $z=\half\cos\theta$. Calculating $g_{\mu\nu}=4{\rm tr} 
(\partial_\mu {\hat\xi} \partial_\nu {\hat\xi})$, we obtain the 
usual spherical metric for the space of pure states. For a 
mixed state we write $(\rho^1,\rho^2,\rho^3)=(u,\theta,\phi)$, where 
$t=(\half-u^2)^{1/2}$, $x=u\sin\theta\cos\phi$,  $y=u\sin\theta \sin\phi$, 
and  $z=u\cos\theta$. Calculating $G_{ab}=4{\rm tr} (\partial_a {\hat\xi} 
\partial_b {\hat\xi})$ we obtain the usual hyperbolic metric for a 
ball of radius $\half$ (interior of the Bloch sphere). 

The richness of the geometry of the space of density matrices, even 
in the $2\times2$ example, is elucidated further by the following 
analysis. We square (\ref{eq:5}) to obtain 
\begin{eqnarray}
{\hat \rho} = \! \left( \!\! 
\begin{array}{cc} t^2+x^2+y^2+z^2+2tz & 2t(x-{\rm i}y) \\
2t(x+{\rm i}y) & t^2+x^2+y^2+z^2-2tz  \end{array} \!\! \right) . 
\nonumber \\
\label{eq:6}
\end{eqnarray}
By construction ${\hat\rho}$ is positive; thus we need only 
consider the trace condition 
\begin{eqnarray}
{\rm tr}({\hat\rho}) = 1\quad  \Longrightarrow \quad t^2+x^2+
y^2+z^2 = \half. \label{eq:7}
\end{eqnarray}
Therefore, under the Hermitian square-root map the space of $2\times2$ 
density matrices is mapped to a sphere $S^{3}\subset 
{\mathds R}^4$. Making use of the trace condition (\ref{eq:7}) 
we can write 
\begin{eqnarray}
{\hat \rho} = \left( \begin{array}{cc} \frac{1}{2}+2tz & 2tx-{\rm i}(2ty) \\
2tx+{\rm i}(2ty) & \frac{1}{2}-2tz \end{array} \right),  \label{eq:8}
\end{eqnarray}
where $t$ is determined by $(x,y,z)$ via (\ref{eq:7}). Clearly, under 
the reflection $(t,x,y,z) \to (-t,-x,-y,-z)$ we have ${\hat\rho} \to 
{\hat\rho}$. This, however, is not the only degeneracy. While a 
density matrix 
has a unique positive semidefinite square root ${\hat\xi}$ and a 
unique negative semidefinite square root $-{\hat\xi}$, there also 
exist square roots which are neither positive nor negative 
semidefinite. 

To work out the roots, suppose that a density matrix 
\begin{eqnarray}
{\hat \varrho} = \left( \begin{array}{cc} \frac{1}{2}+a & b-{\rm i}c \\
b+{\rm i}c & \frac{1}{2}-a \end{array} \right) \label{eq:10}
\end{eqnarray}
is given.
How many points on $S^3$ give rise to the same  
${\hat \varrho}$ of (\ref{eq:10})? By comparing (\ref{eq:8}) and 
(\ref{eq:10}) we obtain: 
\begin{eqnarray}
2tx = b, \quad 2ty = c, \quad 2tz = a. \label{eq:48}
\end{eqnarray}
To find the intersection of (\ref{eq:7}) and (\ref{eq:48}) we 
eliminate $(x,y,z)$ using (\ref{eq:48}) to obtain 
$4t^4 - 2t^2 + R^2 = 0$, 
where $R^2\equiv a^2+b^2+c^2$ (clearly, $0\leq R^2 \leq 
\frac{1}{4}$). This quartic equation admits in general 
four solutions: 
\begin{eqnarray}
t = \pm \sqrt{\frac{1\pm\sqrt{1-4R^2}}{4}} . \label{eq:50}
\end{eqnarray}
We see, therefore, that the space of an equivalence class on 
$S^3$ is already quite nontrivial. 

\begin{figure}[t]
\begin{center}
  \includegraphics[scale=0.2]{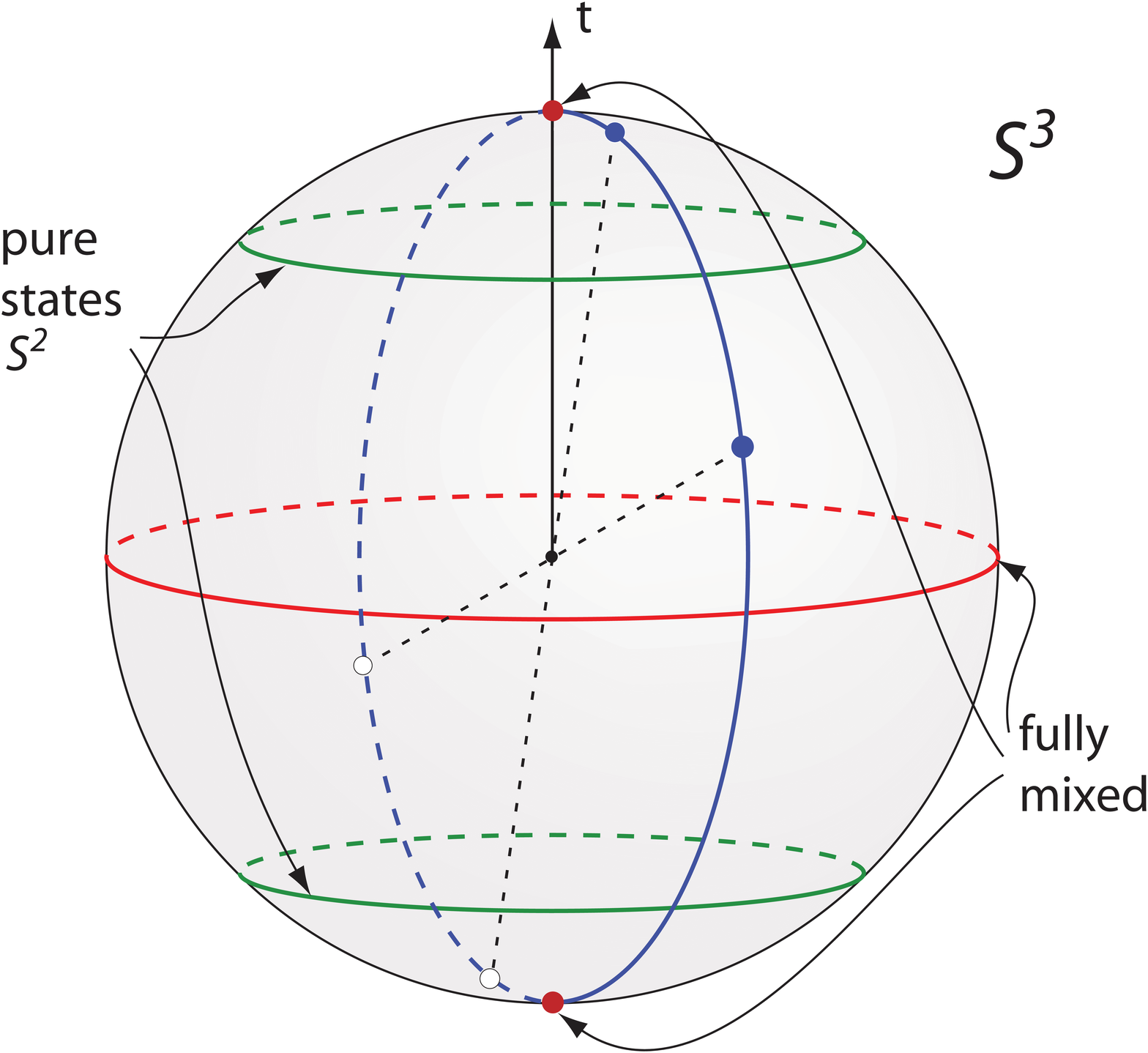}
  \caption{(colour online) 
  \textit{Manifold of $2\times2$ density matrices}. 
  The sphere $S^3$ of radius $1/\surd{2}$ constitutes the covering 
space for the space of $2\times2$ density matrices. If we regard 
the $t$-axis as defining the `poles', then the density matrix 
${\hat \varrho}=\half{\mathds 1}$ corresponds to the `north' and 
the `south' poles, as well as the equator. Moving down from the 
north pole, at $t=\pm\frac{1}{2}$ we have the latitudinal `circles' $S^2$ 
given by $x^2+y^2+z^2 = \quat$, corresponding to the pure-state 
boundary. Crossing this boundary by further reducing the value of 
$t$ we see a `mirror image', where starting from pure state one 
approaches the fully-mixed state ${\hat \varrho}=\half{\mathds 1}$ 
as $t\to0$. The northern hemisphere of the $S^3$ is thus 
partitioned into two parts; the northern cap $t\geq\frac{1}{2}$ and 
the latitudinal band $0\leq t<\frac{1}{2}$. Each point 
is identified with an antipodal point in the southern 
hemisphere, but in addition the points above and 
below $t=\frac{1}{2}$ are also identified in pairs; and finally, the 
equator is identified as a single point, also to be identified 
with the two poles. 
  \label{fig:1} 
  }
\end{center}
\end{figure}

There are three distinct cases: (i) $R=0$ (so that 
${\hat \varrho}$ is `fully mixed', i.e. proportional to the identity); (ii) 
$R=\frac{1}{2}$ (so that ${\hat \varrho}$ is `pure'); and (iii) $0<R<
\frac{1}{2}$ (so that ${\hat \varrho}$ is `generic'). 
In case (i), we have 
either $t=\pm1/\sqrt{2}$ or $t=0$ (double root). For $t=\pm1/\sqrt{2}$ 
we have an antipodal pair on $S^3$, whose coordinates are 
$(t,x,y,z) = (\frac{1}{\surd{2}},0,0,0)$ and $(-\frac{1}{\surd{2}},0,0,0)$,
that give rise to the same density matrix; for $t=0$ we 
have an $S^2$ degrees of freedom 
given by $x^2+y^2+z^2=\frac{1}{2}$; all the points on  this $S^2$ 
determine the density matrix proportional to the identity. In case 
(ii), we have $t=\pm\frac{1}{2}$, and hence 
$(t,x,y,z) = \pm (\frac{1}{2}, b, c, a)$. 
The totality of pure states on $S^3$ is then given by a pair of 
two-spheres $S^2$ associated with $t=\half, x^2+y^2+z^2=\quat$ 
and $t=-\half, x^2+y^2+z^2=\quat$. In case (iii), the generic case, 
we have four distinct values for $t$ given by (\ref{eq:50}). That is, 
two antipodal 
pairs on $S^3$ give rise to the same density matrix (\ref{eq:10}). 
Therefore, the space of mixed-state density matrix is given by the 
quotient of $S^3$ by the identification of two antipodal pairs. These 
configurations are schematically illustrated in figure \ref{fig:1}. 

The utility of information geometry is not merely confined to the 
study of the structure of quantum state spaces; information 
geometry plays a crucial role in the theory of statistical estimation. 
In the quantum context, typically one has a system characterised 
by a density matrix ${\hat\rho}(\theta)$ that depends on a family of 
unknown parameters $\{\theta^i\}$. To determine the state of the 
system one constructs a family of observables $\{{\hat\Theta}^i\}$, 
corresponding to the estimators of $\{\theta^i\}$; by measuring these 
observables one estimates the unknown parameters. The degrees 
of error in these estimates are characterised by the covariance 
matrix ${\rm tr}[({\hat\Theta}^i-\theta^i)({\hat\Theta}^j-\theta^j)
{\hat\rho}(\theta)]$, 
where for simplicity we have assumed that the estimators 
$\{{\hat\Theta}^i\}$ are unbiased so that ${\rm tr}({\hat\Theta}^i 
{\hat\rho}(\theta))=\theta^i$. The lower bound of the covariance 
matrix is determined by the inverse of the quantum Fisher-Rao 
metric $g_{ij} = 4{\rm tr}(\partial_i{\hat\xi}(\theta) 
\partial_j{\hat\xi}(\theta))$. In many applications the trace can be 
computed explicitly to determine the lower bounds for the 
estimates. 

As a concrete example, let us work out the variance lower bound 
associated with the estimation of unitary (e.g., time) parameter. 
The problem is as follows. We have a 
system prepared in an initial state ${\hat\xi}_0={\hat\xi}(0)$, which 
evolves under the Hamiltonian ${\hat H}$ so 
that 
\begin{eqnarray}
{\hat\xi}_t = \exp(-\ri{\hat H}t) {\hat\xi}_0\exp(\ri{\hat H}t).
\end{eqnarray} 
The task is to estimate how much time has elapsed since 
its initial preparation. Let ${\hat T}$ be an unbiased estimator for 
the time parameter $t$ so that ${\rm tr}({\hat T}{\hat\xi}_t
{\hat\xi}_t)=t$ (strictly speaking, all we need is a locally unbiased 
estimator satisfying $\partial_t{\rm tr}({\hat T}{\hat\xi}_t{\hat\xi}_t)
=1$) \cite{Holevo}. 
If we write ${\tilde T}={\hat T}-t{\mathds 1}$ for the mean-adjusted 
estimator, then the variance of ${\hat T}$ is $\Delta T^2
= {\rm tr} ({\tilde T}^2{\hat\xi}_t{\hat\xi}_t)$. 
By differentiating the trace condition ${\rm tr}({\hat\xi}_t{\hat\xi}_t)=1$ 
in $t$ and writing ${\hat\xi}'_t= \partial_t {\hat\xi}_t$, we obtain 
${\rm tr}({\hat\xi}'_t {\hat\xi}_t)=0$. Hence by differentiating 
${\rm tr}({\tilde T}\xi_t\xi_t)=0$ in $t$ and using ${\tilde T}'=
-{\mathds 1}$ we find
\begin{eqnarray}
{\rm tr} [({\tilde T}{\hat\xi}_t+{\hat\xi}_t{\tilde T}){\hat\xi}'_t]=1. 
\label{eq:77} 
\end{eqnarray}
Recall that the matrix Schwarz inequality applied to a pair of 
Hermitian operators ${\hat X}$ and ${\hat Y}$ states that 
$[{\rm tr}({\hat X}{\hat Y})]^2 \leq {\rm tr}({\hat X}^2)
{\rm tr}({\hat Y}^2)$. Squaring (\ref{eq:77}) and 
substituting ${\hat X}={\tilde T}{\hat\xi}_t+{\hat\xi}_t{\tilde T}$ and 
${\hat Y}={\hat\xi}'_t$ in the Schwarz inequality, we deduce a version 
of the quantum Cram\'er-Rao inequality
\begin{eqnarray}
\Delta T^2 + \delta T^2 \geq \frac{1}{2\, 
{\rm tr}({\hat\xi}'_t{\hat\xi}'_t)}, 
\label{eq:17}
\end{eqnarray}
where $\delta T^2 = {\rm tr}({\hat T}{\hat\xi}_t{\hat T}{\hat\xi}_t)
-t^2$, and where the denominator on the right side is half the 
Fisher-Rao metric. By use of the unitary evolution ${\hat\xi}'_t = 
-\ri({\hat H}{\hat\xi}_t-{\hat\xi}_t{\hat H})$ we find that 
\begin{eqnarray}
{\rm tr}({\hat\xi}'_t{\hat\xi}'_t) = 2 \left[ {\rm tr}({\hat H}^2 
{\hat\xi}^2_t) - {\rm tr}({\hat H}{\hat\xi}_t{\hat H}{\hat\xi}_t)
\right]. \label{eq:18}
\end{eqnarray}
The right side of (\ref{eq:18}) is twice the skew 
information introduced by Wigner and Yanase \cite{WY}. It is 
interesting that Fisher introduced the left side of (\ref{eq:18}), 
in the \textit{classical} context, as the amount of extractable 
information concerning the value of $t$ 
\cite{Fisher}, whereas Wigner and Yanase introduced the right 
side of (\ref{eq:18}), in the \textit{quantum} context, as an 
\textit{a priori} measure of information contained in the state 
concerning ``not easily measured quantities'' (such as 
${\hat T}$) \cite{WY}. In special cases for which ${\hat\xi}_t$ 
is a pure state satisfying ${\hat\xi}_t^2={\hat\xi}_t$, the skew 
information is maximised and (\ref{eq:18}) coincides with twice 
the energy variance 
$2\Delta H^2$, while $\delta T^2=0$, and we recover the 
uncertainty relation for the unitary parameter estimation 
as in \cite{Brody1}. For mixed states, however, we obtain 
\begin{eqnarray}
\Delta T^2 + \delta T^2 \geq \frac{1}{4(\Delta H^2 - \delta H^2)} . 
\label{eq:x15}
\end{eqnarray} 

We note that for an arbitrary observable ${\hat H}$ the 
Wigner-Yanase skew information $I_\rho$ is defined by the 
expression
\begin{eqnarray}
I_\rho(H) = {\rm tr}({\hat H}^2{\hat\rho}) - {\rm tr}({\hat H}
\sqrt{\hat\rho} {\hat H}\sqrt{\hat\rho}) , 
\end{eqnarray}
whereas the quantity $\delta H$ introduced above, which might 
appropriately be called the skew information of the second kind, 
is given by 
\begin{eqnarray}
\delta H^2 = {\rm tr}({\hat H}\sqrt{\hat\rho}{\hat H}\sqrt{\hat\rho}) - 
\left[{\rm tr}({\hat H}{\hat\rho})\right]^2 . 
\end{eqnarray}
It follows that the total variance takes the form $\Delta H^2=I_\rho(H)
+\delta H^2$, which can be interpreted as the quantum analogue of the 
conditional variance formula (that the total variance is the sum of the 
expectation of the conditional variance and the variance of the conditional 
expectation). In particular, it is easily 
shown that $\Delta T^2 \geq 
\delta T^2 \geq 0$. By taking the upper bound for $\delta T^2$, 
one obtains 
$\Delta T^2\geq 1/4\, {\rm tr}({\hat\xi}'_t{\hat\xi}'_t)$, which is 
the uncertainty relation based on the skew information obtained 
by Luo \cite{Luo} (see also \cite{Petz2}). This bound is not 
tight because the Luo relation gives $\Delta T^2 
\Delta H^2 \geq 1/8$ for pure states, whereas (\ref{eq:17}) 
gives $\Delta T^2 \Delta H^2 \geq 1/4$. If we combine (\ref{eq:x15}) 
with its dual inequality $\Delta H^2+\delta H^2\geq 1/4(\Delta T^2-
\delta T^2)$, we obtain a less tight but more symmetric expression 
\begin{eqnarray}
\Delta T^2 \Delta H^2 \geq \quat + \delta T^2 \delta H^2 , 
 \label{eq:x18}
\end{eqnarray}
which implicitly appears in \cite{Luo2}.

By exploiting the utility of information geometry, we can go beyond 
the lowest-order term in uncertainty relations. To this end, we 
remark that the Fisher-Rao metric that determines the 
uncertainty lower bound is just the squared `velocity' of the 
evolution of quantum states. In the case of a pure state, the 
fact that squared velocity of the state evolution is given by the 
energy uncertainty is known as the Anandan-Aharonov relation 
\cite{AA}. The observation of Luo can be paraphrased by saying that 
this velocity, in the 
case of a generic mixed state, is given by the skew information. 
In addition to the squared velocity, we consider the squared 
`acceleration' ${\hat\alpha}_t = {\hat\xi}''_t - {\rm tr}({\hat\xi}''_t 
{\hat\xi}_t) {\hat\xi}_t$, which determines the curvature 
$\gamma^2=16 {\rm tr}({\hat\alpha}_t{\hat\alpha}_t)/
{\rm tr}({\hat\xi}'_t{\hat\xi}'_t)$ of the curve in the space of 
square-root density matrices. The squared acceleration 
determines the next-order correction to the uncertainty 
relation. More generally, we can obtain arbitrarily many 
higher-order corrections by considering higher-order derivative 
terms, thus generalising the quantum Bhattacharrya bounds 
derived in \cite{Brody1,Brody2,Brody3} for pure states to general 
density matrices. 

To illustrate how higher-order corrections to uncertainty relations 
can be obtained systematically, let us examine the second-order 
and the third-order corrections here. For this purpose, it will be 
convenient to relax the unit trace condition ${\rm tr}{\hat\rho}=1$ 
but merely require ${\rm tr}{\hat\rho}\neq0$ and ${\rm tr}{\hat\rho}
<\infty$, and impose unit trace at the end of the calculation. 
We then define the symmetric function $t({\hat\xi})={\rm tr}(
{\hat\xi}{\hat T}{\hat\xi})/{\rm tr}({\hat\xi}{\hat\xi})$ of ${\hat\xi}$ and 
consider the gradient $\nabla t$ of this function in the space 
of square-root density matrices, evaluated at ${\rm tr}({\hat\xi}
{\hat\xi})=1$. A short calculation then shows that the squared 
magnitude (in the Hilbert-Schmidt norm) of the gradient is 
$|\nabla t|^2 = 2(\Delta T^2 + \delta T^2)$. On the other hand, 
the squared magnitude of the gradient is clearly larger than 
(or equal to) the squared magnitude of its component in the 
direction of ${\hat\xi}'_t$, and thus 
$\Delta T^2 + \delta T^2 \geq 1/2{\rm tr}({\hat\xi}'_t{\hat\xi}'_t)$,
where we have made use of the relation 
${\rm i}[{\hat H},{\hat T}]=1$. 
This provides an alternative derivation 
of the quantum Cram\'er-Rao inequality. 

The next order correction to the uncertainty lower bound for 
mixed states is obtained from the component of $\nabla t$ in 
the direction of ${\hat\alpha}_t$, given by 
${\rm tr}(({\hat T}{\hat\xi}_t+{\hat\xi}_t{\hat T}){\hat\alpha}_t)/
{\rm tr}({\hat\alpha}_t{\hat\alpha}_t)$. Hence if the curvature 
$\gamma$ of the curve ${\hat\xi}_t$ is 
large, the correction is small. The numerator, 
however, in general depends on ${\hat T}$. To 
seek a bound that is independent 
of ${\hat T}$ we thus consider the third-order term, i.e. 
the component of the gradient in the direction of ${\hat\beta}_t 
= {\hat\xi}_t'''-({\rm tr}({\hat\xi}_t'''{\hat\xi}_t')/
{\rm tr}({\hat\xi}_t'{\hat\xi}_t')){\hat\xi}_t'$. This is given by 
${\rm tr}(({\hat T}{\hat\xi}_t+{\hat\xi}_t{\hat T}){\hat\beta}_t)/
{\rm tr}({\hat\beta}_t{\hat\beta}_t)$. Although the calculation of 
this term is straightforward, the resulting 
expression is lengthy and will be omitted here. It suffices to 
note that each of the odd-order corrections involves commutators
between ${\hat T}$ and ${\hat H}^k$ for some $k$, and 
thus is independent of ${\hat T}$; whereas all 
even-order corrections are in general dependent on ${\hat T}$. 
Furthermore, terms appearing in the corrections have 
specific statistical interpretations in terms of the quantum extensions 
of central moments of the Hamiltonian. In the pure state 
limit where ${\hat\xi}_t^2={\hat\xi}_t$, these expressions reduce 
to the ``classical'' central moments, in a manner analogous to 
the reduction of the Wigner-Yanase skew information to the 
second central moment $\Delta H^2$. We are thus able to 
obtain various higher-order generalisations of the skew 
information, which appear to be entirely new quantities in the 
study of statistical properties of mixed states in quantum 
mechanics. These quantities might appropriately be called 
\textit{quantum skew moments}.

\vskip 4pt I thank E.~J.~Brody, E.~M.~Graefe, L.~P.~Hughston, 
M.~F.~Parry, and the participants of the third international 
conference on Information Geometry and Its Applications, Leipzig 
2010, for stimulating discussions. 

\vskip -8pt



\end{document}